# GPU based Parallel Genetic Algorithm for Solving an Energy Efficient Dynamic Flexible Flow Shop Scheduling Problem

Jia Luo[a,*], Shigeru Fujimura[b], Didier El Baz[a]

a LAAS-CNRS, Université de Toulouse, CNRS, Toulouse, France

b Graduate School of Information, Production, and Systems, Waseda University, Kitakyushu, Japan

## Abstract

Due to new government legislation, customers' environmental concerns and continuously rising cost of energy, energy efficiency is becoming an essential parameter of industrial manufacturing processes in recent years. Most efforts considering energy issues in scheduling problems have focused on static scheduling. But in fact, scheduling problems are dynamic in the real world with uncertain new arrival jobs after the execution time. This paper proposes a dynamic energy efficient flexible flow shop scheduling model using peak power value with the consideration of new arrival jobs. As the problem is strongly NP-hard, a priority based hybrid parallel Genetic Algorithm with a predictive reactive complete rescheduling approach is developed. In order to achieve a speedup to meet the short response in the dynamic environment, the proposed method is designed to be highly consistent with NVIDIA CUDA software model. Finally, numerical experiments are conducted and show that our approach can not only achieve better performance than the traditional static approach, but also gain competitive results by reducing the time requirements dramatically.

---

[*] Corresponding author.

# 1. Introduction

About one half of the world's total energy is currently consumed by the industrial sector [1] and its energy consumption has nearly doubled over the last 60 years [2]. Thus energy efficiency is becoming an essential parameter of industrial manufacturing processes, mostly due to new government legislation, customers' environmental concerns and continuously rising cost of energy. Because of a growing economical competitive landscape and higher environmental norms, it is now vital for manufacturing companies to reduce their energy consumption and to become more environment-friendly.

The adjustment of scheduling strategies only requires modest time and cost investment, compared with the redesign methods for machines or processes [3]. Therefore, a lot of traditional scheduling strategies considering minimizing the total energy consumption have been studied [4, 5, 6]. Meanwhile, some efforts have been made on taking peak power into account, because electricity consumption and operating costs of manufacturing plants are usually charged based on peak power demand from electricity providers [7]. However, most of the research works only concentrate on establishing a mathematical model for solving the optimization problem in a static environment. But in fact, scheduling problems are dynamic in the real world with uncertain new arrival jobs after the start time. Few works take [8, 9] reactive approaches into consideration for supporting energy efficient dynamic systems. Moreover, they care only about the improvement of algorithms to gain better solution quality, while ignoring the time consumption of the implementation of such approaches. Without a doubt, a method proposing an adequate rescheduling plan in a short response time is greatly desired in this case.

In the last decade, Graphics Processing Units (GPUs) have gained widespread popularity as computing accelerators for High Performance Computing (HPC) applications [10]. Research on GPU-based approaches for solving scheduling problems [11, 12, 13, 14] has won favor in recent years with the development in 2006 of Compute Unified Device Architecture, CUDA (a software and hardware architecture that enables GPUs to be programmed with some high level programming languages like C, C++ and Fortran) [15]. Despite all those advances, the complex problem as dynamic energy efficient flexible flow shop scheduling has not been considered as far our knowledge is concerned. Additionally, many parallel Genetic Algorithm (GA) implementations for solving optimization problems have shown their success [16, 17, 18] as seen in the literature. Therefore, the GPU based parallel GA for solving an energy efficient dynamic flexible flow shop scheduling problem remains an open research challenge based on the previous works, and the one that we seek to address in this paper.

The total tardiness and the makespan with a peak power limitation are analyzed in this paper while considering a dynamic environment in the flexible flow shop. A predictive reactive complete rescheduling approach is adopted to represent the optimization problem. Furthermore, due to the fact that an adequate renewed scheduling plan needs to be obtained in a short response time in the dynamic environment, a priority based parallel GA on GPUs is implemented. The efficiency and the effectiveness of the proposed approach are validated through computational tests. Specially, the contributions of our work are summarized as followed:

1. We propose a dynamic energy efficient flexible flow shop scheduling model using peak power value with the consideration of new arrival jobs.

2. A priority based hybrid parallel GA mapping to NVIDIA CUDA software model is developed with a predictive reactive complete rescheduling approach.

3. Our method can not only achieve better performance than the traditional static approach, but also gain competitive results by reducing the time requirements dramatically.

The remaining sections of this paper are organized as follows. Section 2 introduces related works. Section 3 describes the research problem and the mathematical model. Section 4 presents the priority based hybrid parallel GA for solving the energy efficient dynamic flexible flow shop scheduling problem. Section 5 illustrates the numerical experiments and result analysis. Finally, section 6 states the conclusions.

## 2. Related Works

Recently, there has been growing interest in reducing the energy consumption in manufacturing processes. Several works tried to reduce the peak power in parallel multi-machine contexts. Fang et al. [4] presented a multi-objectives mixed-integer programming model of the flow shop scheduling problem that considers peak power load, energy consumption, and associated carbon footprint in addition to cycle time. Bruzzone et al. [19] proposed the integration of an energy aware scheduling module, with an advanced planning and scheduling system in order to control the peak consumption, while accepting a possible increase in the total tardiness. Xu et al. built a discrete-time mixed-integer programming model and a slot-based mixed-integer programming model in [7] to achieve a global optimal solution between the peak power and the traditional production efficiency without any compromise on computing efficiency. As delaying the production activities may not be acceptable in manufacturing, minimizing the total energy consumption within the traditional scheduling problem is an alternative solution. Liu et al. [5] developed a model for the bi-objectives problem that minimizes total electricity consumption and total weighted tardiness, where the non-dominant sorting genetic algorithm is employed to obtain the Pareto front. Similarly, an emission-aware multi-machine job shop scheduling model was addressed in [20] and solved through a modified multi-objectives genetic algorithm. Dai et al. [21] reported an energy efficient model for the flexible flow shop scheduling and employed a genetic-simulated annealing algorithm to make a significant tradeoff between the makespan and the total energy consumption. To sum up, numerous works have focused on energy efficient scheduling for various shop floor environments in static perspective. But, due to frequently inevitable new arrival jobs in the production environment, a fixed preset scheduling plan could not meet the requirement.

Dynamic scheduling problems are more complex than static scheduling problems. A lot of methods have been hired to solve this kind of problems [22]. Most of them only considered the efficiency of the traditional scheduling problem without including energy efficient demand. Tang et al. [9] adopted a predictive reactive approach based on an improved particle swarm optimization to search for the Pareto optimal solution in dynamic flexible flow shop scheduling problems reducing the energy consumption and the makespan. Pach et al. [23] set up a potential fields based reactive scheduling approach for flexible manufacturing systems in which resources are able to switch to standby mode to avoid useless energy consumption and to emit fields to attract products. Zhang et al. [8] proposed a goal programming mathematical model, which considers the energy consumption and the scheduling efficiency simultaneously to solve the dynamic scheduling problem in flexible manufacturing systems. In a word, some efforts to solve energy efficient dynamic scheduling problems have been carried out. However, some limitations still remain that must be tackled. A typical one is to obtain the renewed adequate scheduling plan in a reasonable response time, particularly for large-scale manufacturing problems.

In recent years, various algorithms, like branch and bound, genetic algorithms, Tabu search, using GPUs have been successfully employed to generate optimized results for scheduling problems with impressive time decrease. Melab et al. [12] indicated a parallel branch and bound algorithm based on a GPU-accelerated bounding model on flow shop scheduling benchmarks to improve the performance by optimizing data access management. Czapinski et al. [24] implemented a Tabu search meta heuristic method on GPUs for the solution of the permutation flow shop scheduling problem, which gains 89 times faster than the CPU version. Zajicek et al. [25] studied a parallel island-based genetic algorithm for the solution of the flow shop scheduling problem by carrying out all computations on GPUs in order to reduce communication. Pinel et al. [13] presented GPU implementations on the Min-Min heuristic and the GraphCell, an advanced parallel cellular genetic algorithm, for solving large instances of the scheduling of independent tasks problem. An improved genetic algorithm and its implementation on CUDA to search optimal solutions to the flow shop scheduling problems with fuzzy processing times and fuzzy due dates were discussed in [26]. These cases have confirmed that the parallel GA on GPUs has good performance in solving scheduling problems. However, it is also revealed that few studies have been conducted to integrate GPUs in dynamic energy efficient scheduling problems, because of the complexity that is caused.

Although many research works on scheduling problems have been studied in GPU literatures, none of them have so far, and to the best of our knowledge, considered energy saving strategies and dynamic environment completely. The above-mentioned efforts provide a starting point for exploring the GPU based parallel GA for solving an energy efficient dynamic flexible flow shop scheduling problem with competitive results and dramatical time reduction.

## 3. Problem statement

3.1 EDFFS problem description

The flexible flow shop scheduling problem (FFS) is a multistage production process that consists of two or more stages in series as illustrated in Fig. 1. There is at least one machine in each stage, and at least one stage has more than one machine. All the jobs need to go through all the stages in the same order before they are completed. On each stage, one machine is selected for processing a given operation. An energy efficient dynamic flexible flow shop scheduling (EDFFS) is a further development of the FFS. A set of new jobs may arrive after the start of the original plan. They should be processed sequentially and non-preemptively from the beginning of the rescheduling point with the remaining uncompleted operations of the original jobs.

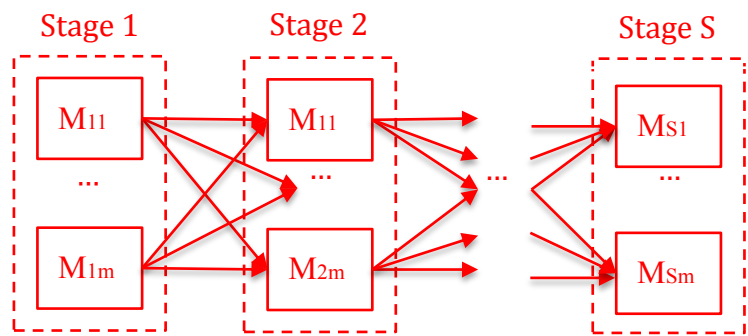

**Fig.1.** A flexible flow shop layout

One instance of the EDFFS problem consists of a set of J jobs and a set of M machines. Each job j∈J on machine m∈M has a corresponding processing time and power consumption. As an FFS problem is considered to be NP-hard in essence and difficult to solve [27], the EDFFS problem is a NP-hard combinatorial optimization problem and more complex than the FFS problem. Additionally, required conditions for the EDFFS are shown in Table 1.

**Table 1**

Required conditions for EDFFS.

| Number | Description |
|---|---|
| 1 | Each operation of a job must be processed by one and only one machine. |
| 2 | Each machine can process no more than one operation at a time. |
| 3 | There is no precedence between operations of different jobs, but there is precedence among operations due to the jobs' processing cycles. |
| 4 | Preemptive operations are not allowed. |
| 5 | Each job is available for processing after the release time. |
| 6 | Machines may suffer new arrival jobs at any time after the rescheduling point. |
| 7 | Processing times and average power consumption for any operation of all jobs on any machine are known. |
| 8 | Setup times for job processing and machine assignment times between stages are not taken into consideration. |
| 9 | There is infinite intermediate storage between machines. |

3.2 Mathematical model of EDFFS

For an easy presentation, we summarize the notations used along the rest of the paper in Table 2.

**Table 2**

A description of notations used in all formulae.

| Notation | Description |
|---|---|
| $j, i, i'$ | Job indices |
| $s, s', s''$ | Stage indices |
| $m$ | Machine index |
| $t$ | Time period index |
| $n$ | Number of original jobs |
| $n'$ | Number of new arrival jobs |
| $r$ | Number of original jobs assigned to machines before the rescheduling point |
| $g$ | Number of stages |

| | | |
|---|---|---|
| o | Number of machines at the stage s. Each stage has the same amount of machines | |
| H | Time horizon | |
| J | Set of original jobs, $J = \{0,1,2,\ldots,n-1\}$ | |
| J' | Set of new arrival jobs, $J' = \{0,1,2,\ldots,n'-1\}$ | |
| S | Set of stages, $S = \{0,1,2,\ldots,g-1\}$ | |
| M | Set of machines at the stage, $M = \{0,1,2,\ldots,o-1\}$ | |
| T | Set of time periods, $T = \{1,2,3,\ldots,H\}$ | |
| RS | Rescheduling point | |
| $R_j$ | Release time of job j, $j \in J \cup J'$ | |
| $D_j$ | Due time of job j, $j \in J \cup J'$ | |
| $P_{jsm}$ | Processing time when job j at stage s is to be processed on machine m, $j \in J \cup J', s \in S, m \in M$ | |
| $Q_{jsm}$ | Average power consumption when job j at stage s is to be processed on machine m, $j \in J \cup J', s \in S, m \in M$ | |
| $Q_{max}$ | Power's peak | |
| WT | Weight for the total tardiness in the objective function | |
| $u_{jst}$ | Boolean variable, $j \in J \cup J', s \in S, t \in T$ | |
| $S_{js}$ | Start time of job j at stage s, $j \in J \cup J', s \in S$ | |
| $M_{js}$ | Target machine handling job j at stage s, $j \in J \cup J', s \in S$ | |
| $Q_t$ | Total power consumption at time period t, $t \in T$ | |
| $T_j$ | Total tardiness, $j \in J \cup J'$ | |
| $C_{max}$ | Completion time of the last job, i.e., the makespan | |
| k | Current generation number of the GA | |
| X(k) | Target machine matrix at generation k | |
| Y(k) | Priority matrix at generation k | |
| Z(k) | Order matrix at generation k | |
| C | A very large constant, $C \in R_+$ | |

To achieve the power's peak limitation and minimize the traditional makespan and the total tardiness objective, the formal mathematical model for the EDFFS is an extension of the mathematical model presented in [7, 19] to cover rescheduling. The formulation is given in the following.

Objective function:

$$\min: WT * \sum_{j \in J \cup J'} T_j + C_{max} \tag{1}$$

Constraints:

$$T_j = \max\{S_{j\,g-1} + P_{j\,g-1\,M_{j\,g-1}} - D_j, 0\} \quad j \in J \cup J' \tag{2}$$

$$C_{max} = \max_{j}\{S_{j\,g-1} + P_{j\,g-1\,M_{j\,g-1}}\} \quad j \in J \cup J' \tag{3}$$

$$S_{j0} \geq R_j \quad j \in J \cup J' \tag{4}$$

$$S_{js} \geq S_{j\,s-1} + P_{j\,s-1\,M_{j\,s-1}} \quad j \in J \cup J', s \in S, s > 0 \tag{5}$$

$$S_{js} + P_{jsM_{js}} \leq S_{is} \quad j \in J \cup J', i \in J \cup J', s \in S, j \neq i, M_{js} == M_{is}, S_{js} \leq S_{is} \tag{6}$$

$$Q_{max} \geq Q_t \quad t \in T \tag{7}$$

$$Q_t = \sum_{j \in J \cup J'} \sum_{s \in S} Q_{jsM_{js}} * u_{jst} \quad t \in T \tag{8}$$

$$u_{jst} = \begin{cases} 1 & j \in J \cup J', s \in S, S_{js} \leq t < S_{js} + P_{jsM_{js}} \\ 0 & otherwise \end{cases} \tag{9}$$

$$RS \leq S_{js} \quad j \in J \cup J', s \in S \tag{10}$$

The decision variables in this mathematical model are $M_{js}$ and $S_{js}$. As two scheduling objectives are considered, it is formulated as a single additive objective function (1) by aggregating the total tardiness and the makespan with the weight WT. As tardy jobs typically cause penalty costs [28] and have a great influence on customer satisfaction, the weight WT indicates the priority of the first objective. Constraints (2) and (3) define the tardiness of the jobs and the makespan separately. The precedence among operations due to the jobs' processing cycles is presented by constraints (4) and (5), while constraint (6) establishes the precedence caused by the sequencing on machines. In addition, constraint (7) introduces the power's peak by an upper bound whereas the power consumption during a certain period is expressed by constraint (8). Constraint (9) gives the definition of a Boolean variable $u_{jst}$. It is equal to 1 if job j at stage s is being processed at time period t. Finally, constraint (10) imposes the definition of rescheduling.

## 4. Solving approach

4.1 Predictive reactive complete rescheduling strategy

The predictive reactive method is the most common dynamic scheduling approach used in manufacturing systems [22]. To solve the EDFFS, operations are assigned to machines in order, following the original schedule until the reschedule point. New arrival jobs and uncompleted operations of original jobs are processed in terms of the updated schedule executed by the optimization algorithm within a short response. A hybrid parallel GA on GPUs is proposed for solving the problem with a complete rescheduling strategy which is better in maintaining optimal solutions, but is rarely achievable in practice due to the prohibitive computation time [22]. Fig.2 summarizes the flow of the predictive reactive complete rescheduling process.

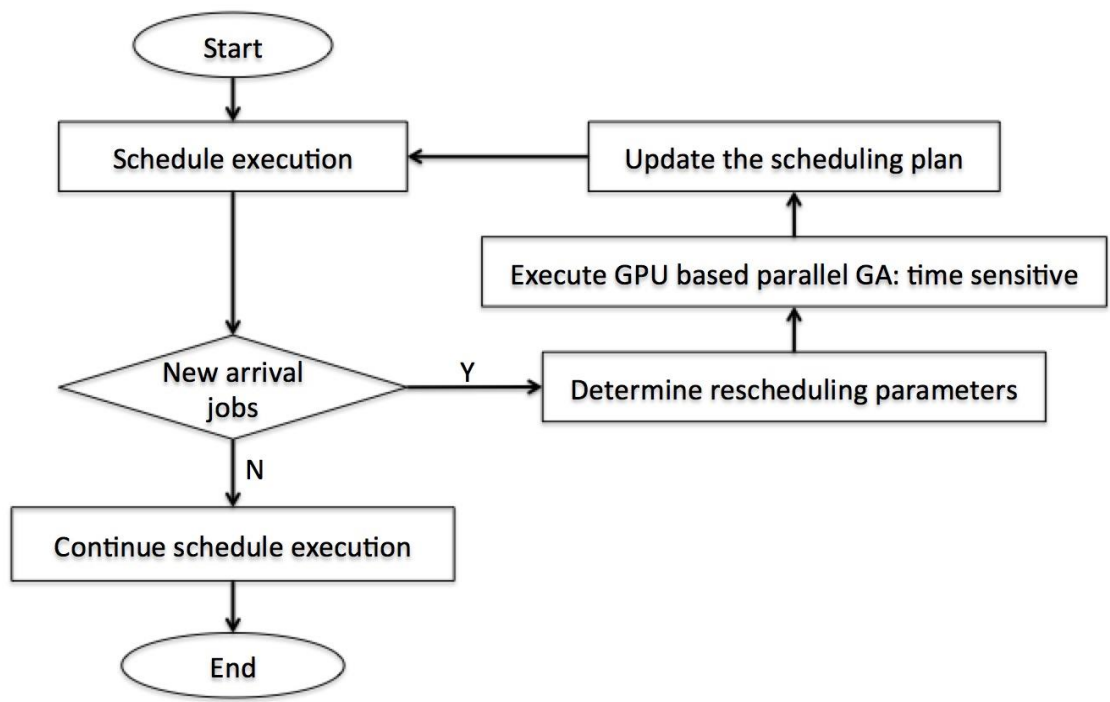

Fig.2. Flow of the predictive reactive complete rescheduling process for the EDFFS

4.2 Hybrid parallel GA model

The Genetic Algorithm (GA) is a stochastic search algorithm based on the principle of natural selection and recombination [29]. However, there is an increase in the required time to find adequate solutions when GA is applied to more complex and larger problems. Parallel implementation is considered as one of the most promising choices to make it faster.

The CUDA framework is chosen to parallelize the GA on GPUs in this paper. It is a Single Instruction, Multiple Threads (SIMT) parallel programming model. The parallel threads are grouped into blocks which are organized in a grid [30] as shown in Fig. 3 using the local memory, the shared memory and the global memory respectively.

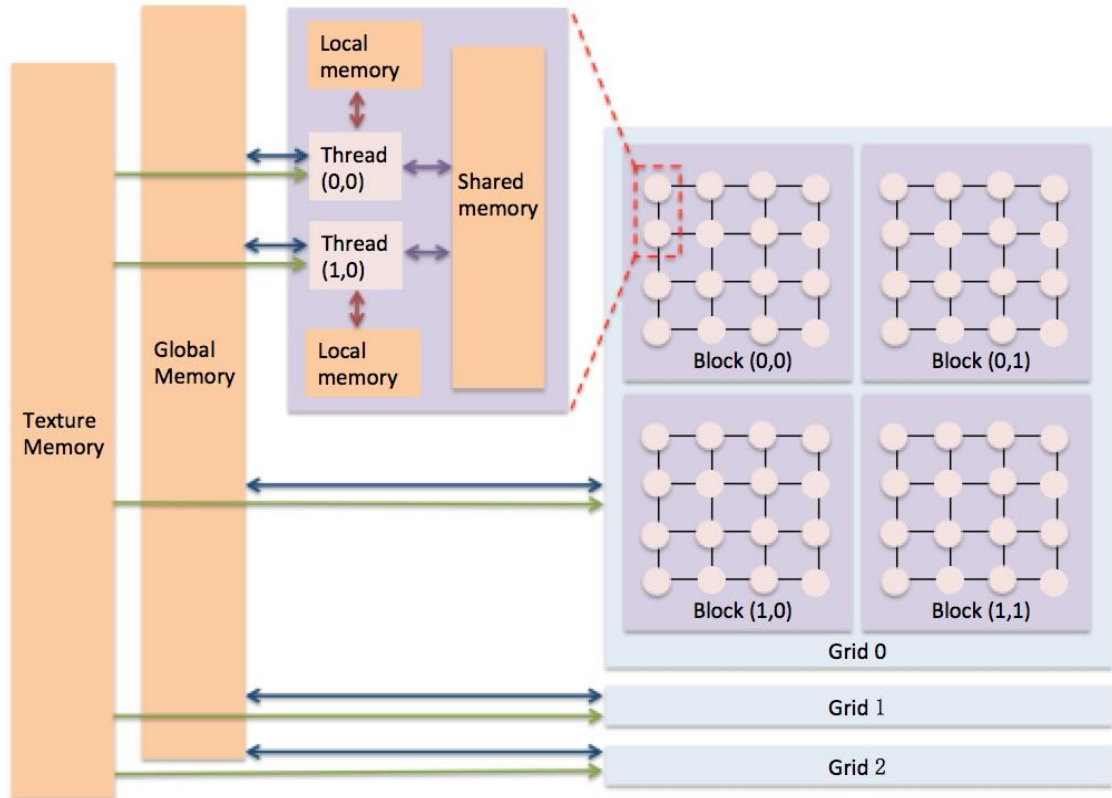

Fig.3. Hierarchy of threads and different types of memory of CUDA framework

There are different ways of exploiting parallelism in GA: master-slave models, fine-grained models, island models, and hybrid models [31]. Fine-grained models can perform well due to the larger genetic diversity obtained by dividing the population into a number of subpopulations [32]. Island models are the most famous for the research on parallel GA. Populations on the islands are free to converge toward different sub-optima with a faster improvement of the average fitness [31] and a migration operator can help mix good features that emerge from the local island. To obtain a good speedup with CUDA and to combine the advantage of fine-grained models and island models, we establish the hybrid model presented in Fig. 4 with a fine-grained GA at the lower level and an island GA at the upper level. A correspondence between the parallel hybrid GA components and the hierarchy of CUDA threads is displayed by Table 3. It turns out that this hierarchy is highly consistent with the hierarchy of threads and different types of memory of CUDA framework.

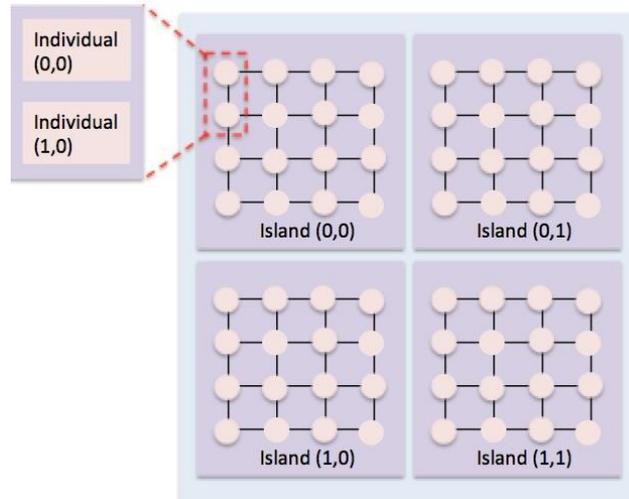

Fig.4. Hierarchy of hybrid GA

**Table 3**

Correspondence between the parallel hybrid GA components and the hierarchy of CUDA threads

| Hybrid GA components | CUDA underlying architecture |
| --- | --- |
| Individual | Thread |
| Island | Block |
| Population | Grid |

At the lower level, each CUDA thread processes one GA individual. Because of the 2D grid, the GA individuals can get connected completely with this topology. A tournament-based selection is executed on texture memory to gain its fast response to read information from neighbors. Crossover, mutation and fitness function calculation are generated using the global memory. On the other hand, one block in CUDA represents one island in GA at the upper level. An elitism based replacement after every generation inside the island and the migration among islands every 10 generations are mainly carried via shared memory. The procedure of the hybrid GA with memory management is expressed in Fig. 5. More details are discussed in section 4.4.

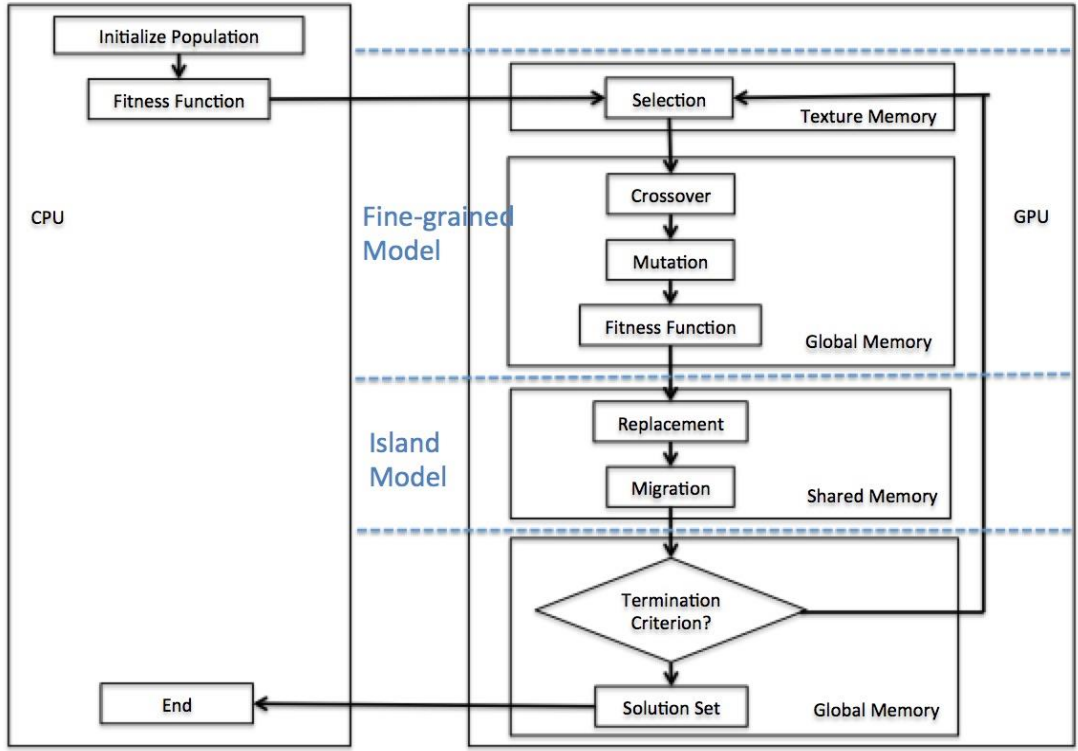

Fig.5. Procedure of the parallel GA with memory management

4.3 Priority based encoding representation

According to the problem description in Section 3, a target machine matrix X(k) stored on GPU global memory with $n + n'$ rows and g columns, is presented in (10).

$$X(k) = \begin{bmatrix} x_{00}(k) & x_{01}(k) & \cdots & x_{0(g-1)}(k) \\ x_{10}(k) & x_{11}(k) & \cdots & x_{1(g-1)}(k) \\ \vdots & \vdots & x_{js}(k) & \vdots \\ x_{(n+n'-1)0}(k) & x_{(n+n'-1)1}(k) & \cdots & x_{(n+n'-1)(g-1)}(k) \end{bmatrix} \quad (10)$$

where $x_{js}(k) \in [0, o-1] \cup \{-1\}, j \in J \cup J', s \in S$.

Moreover, (11) shows a $(n + n') \times g$ matrix placed on GPU global memory that expresses the priority relation among operations.

$$Y(k) = \begin{bmatrix} y_{00}(k) & y_{01}(k) & \cdots & y_{0(g-1)}(k) \\ y_{10}(k) & y_{11}(k) & \cdots & y_{1(g-1)}(k) \\ \vdots & \vdots & y_{js}(k) & \vdots \\ y_{(n+n'-1)0}(k) & y_{(n+n'-1)1}(k) & \cdots & y_{(n+n'-1)(g-1)}(k) \end{bmatrix} \quad (11)$$

where $y_{js}(k) \in [1, g \times (n + n') - r] \cup \{-1\}, j \in J \cup J', s \in S$.

Each element of matrix X(k) indicates the machine number that deals with job j at stage s at generation k while each element of matrix Y(k) is used to sequence the operations assigned to machines. The values for the EDFFS are defined as:

- if job j at stage s is started or completed before the start time of the rescheduling point, both element $x_{js}(k)$ and element $y_{js}(k)$ are equal to -1. This includes:

    Case 1: job j at stage s of the original job is accomplished.
    Case 2: job j at stage s of the original job is being executed.

- if job j at stage s is assigned to a machine after the start time of the rescheduling point, element $x_{js}(k)$ is equal to a random integer representing the target machine handling job j at stage s. Similarly, elements $y_{js}(k)$ is also generated randomly from the range starting from 1 to the amount of unassigned operations. Moreover the value of element $y_{js}(k)$ is unique, where the larger the value of the random integer represents higher priority. This includes:

    Case 1: job j at stage s of the original job remains to be processed.
    Case 2: job j at stage s of the new arrival job must be processed.

In this representation, each chromosome of the parallel GA consists of one target machine matrix and one priority matrix, representing a feasible schedule. In the decoding step, elements of a matrix Z(k) (12) generated from the matrix X(k) and the matrix Y(k) are designed to address the assignment order of uncompleted operations. Element $z_{js}(k)$ is equal to 0 if job j at stage s of the original job is being executed at the start time of the rescheduling point, while element $z_{js}(k)$ is equal to C if the operation is accomplished before it. Elements' value of matrix Z(k) are reserved on GPU global memory and the procedure to determine them is displayed in Algorithm 1. The later assigned operation needs be delayed when the power's peak is met as shown by the decoding rule in Algorithm 2.

$$Z(k) = \begin{bmatrix} z_{00}(k) & z_{01}(k) & \cdots & z_{0(g-1)}(k) \\ z_{10}(k) & z_{11}(k) & \cdots & z_{1(g-1)}(k) \\ \vdots & \vdots & z_{js}(k) & \vdots \\ z_{(n+n'-1)0}(k) & z_{(n+n'-1)1}(k) & \cdots & z_{(n+n'-1)(g-1)}(k) \end{bmatrix} \quad (12)$$

Where $z_{js}(k) \in [1, g \times (n + n') - r] \cup \{0, C\}, j \in J \cup J', s \in S$.

**Algorithm 1**

The procedure for determining elements' value of matrix Z(k)

---

For $s, s', s'' \in S$, $s \neq s''$, $j, i \in J \cup J'$, $j \neq i$, $m \in M$

**if** $x_{js}(k) = -1$ **then**

  **if** $S_{js} < RS < S_{js} + P_{jsm}$ **then**

    $z_{js}(k) = 0$, machine m continues to process job j at stage s before executing a rescheduling plan.

  **else**

    $z_{js}(k) = C.$

  **end if**

**else**

  **if** $y_{js}(k) > y_{is'}(k)$ **then**

    $z_{js}(k) < z_{is'}(k)$, job j at stage s is assigned to its target machine earlier than job i at stage $s'$.

  **end if**

  **if** $s < s''$ **then**

    $z_{js}(k) < z_{js''}(k)$, job j at stage s is assigned to its target machine earlier than job j at stage $s''$.

  **end if**

**end if**

**Algorithm 2**

The decoding rule

---

For $s, s' \in S$, $j, i, i' \in J \cup J'$, $j \neq i \neq i'$, $z_{js}(k) < z_{is}(k)$, $z_{i's'}(k) < z_{js}(k)$ && job $i'$ at stage $s'$ is the earliest finished one among all the processing operations at period t, $m \in M, t \in T$

**if** $Q_{max} \geq Q_t + Q_{jsM_{js}}$ **then**

   **if** $M_{js} == M_{is}$ **then**

      job j at stage s is assigned to machine m earlier than job i at stage s.

   **else**

      jobs are assigned to each machine in terms of matrix X(k).

   **end if**

**else**

   job j at stage s needs be delayed to be assigned to its target machine until finishing job $i'$ at stage $s'$.

**end if**

---

An example of EDFFS is presented in Table 4. There are 6 original jobs. Each job consists of 3 stages and there are two machines at each stage. Jobs are available to be assigned to machines after the release time ($R_j$). Each operation is processed on the target machine ($M_{js}$) after the start time ($S_{js}$). To make it simple, the processing time is set as 1, 2 and 3 for the three stages respectively. The average power consumption $Q_{jsm}$ is defined as 1 for any operation on any machine while the value of the power's peak $Q_{max}$ is equal to 3. Finally, we assign a priority to the total tardiness over the makespan in the objective function by setting WT as 100. Fig. 6 shows the Gantt chart of this scheduling. Regarding new arrival jobs, job 6 and job 7 need be considered after starting the plan. In the traditional static environment, they could only be scheduled after completing the operations of the original schedule at each stage as illustrated in Fig. 7. However, the predictive reactive complete rescheduling approach in a dynamic environment reschedules new arrival jobs at the beginning of the rescheduling point (RS=7) with remaining operations of original jobs simultaneously as in Fig. 8.

**Table 4**

An example of EDFFS

|  | Original jobs | | | | | | New arrival jobs | |
|---|---|---|---|---|---|---|---|---|
|  | job 0 | job 1 | job 2 | job 3 | job 4 | job 5 | job 6 | job 7 |
| $R_j$ | 0.80 | 1.42 | 3.54 | 3.77 | 4.91 | 2.45 | 7.77 | 7.49 |
| $M_{j0}, M_{j1}, M_{j2}$ | 1, 1, 0 | 0, 0, 1 | 0, 1, 1 | 0, 1, 0 | 1, 1, 1 | 0, 0, 0 | | |
| $S_{j0}, S_{j1}, S_{j2}$ | 0.80, 1.80, 3.80 | 1.42, 2.42, 4.42 | 6.80, 9.91, 11.91 | 3.77, 7.91, 12.42 | 4.91, 5.91, 7.91 | 2.45, 7.42, 9.42 | | |

The following matrices show the EDFFS decoding result for the example. Each row of these matrices represents a job and each column represents a stage. A chromosome consists of the target machine matrix X(k) and the priority matrix Y(k) generated randomly to obtain the order matrix Z(k).

$$X(k) = \begin{bmatrix} -1 & -1 & -1 \\ -1 & -1 & -1 \\ -1 & 1 & 0 \\ -1 & 1 & 1 \\ -1 & -1 & 0 \\ -1 & 0 & 0 \\ 1 & 1 & 1 \\ 0 & 0 & 0 \end{bmatrix}, Y(k) = \begin{bmatrix} -1 & -1 & -1 \\ -1 & -1 & -1 \\ -1 & 4 & 2 \\ -1 & 10 & 6 \\ -1 & -1 & 12 \\ -1 & 9 & 3 \\ 8 & 13 & 11 \\ 1 & 5 & 7 \end{bmatrix} \rightarrow Z(k) = \begin{bmatrix} C & C & C \\ C & C & 0 \\ 0 & 8 & 10 \\ C & 2 & 4 \\ C & 0 & 1 \\ C & 3 & 9 \\ 5 & 6 & 7 \\ 11 & 12 & 13 \end{bmatrix}$$

Following the description, the later assigned operation needs be delayed when the power's peak is met. For instance, job 7 at stage 0 was supposed to be processed after the completion of job 2 at stage 0 on machine 0 as in Fig. 8. Moreover, at the same moment machine 2, 3 and 4 are busy with job 5 at stage 1, job 3 at stage 1 and job 4 at stage 2 respectively. But due to power limitation, this scenario is not possible. As $z_{70}(k)$ is equal to 11, $z_{51}(k)$ to 3, $z_{31}(k)$ to 2, $z_{42}(k)$ to 1, job 7 at stage 0 is the newest allocated one among all of them. Thus, it is delayed until the completion of job 5 at stage 2 on machine 4.

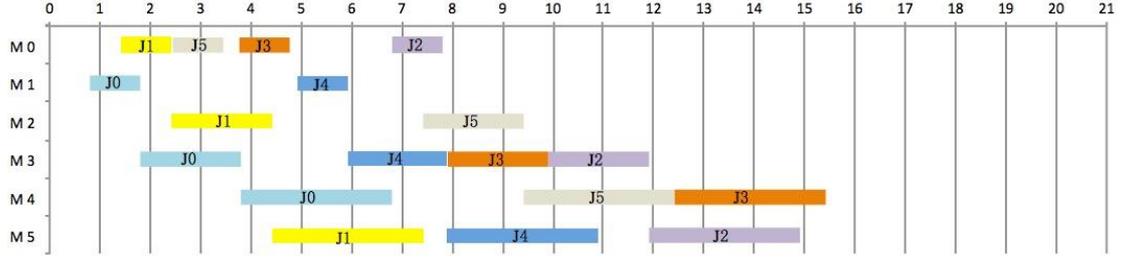

Fig. 6. The original schedule of an optimized solution

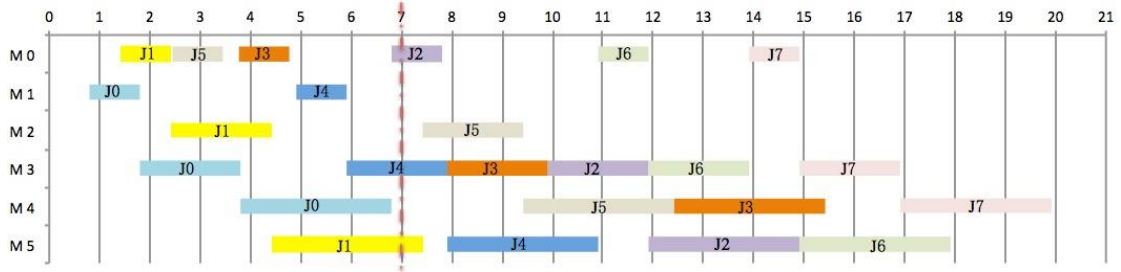

Fig. 7. The updated schedule of an optimized solution in a static environment
( $\sum_{j \in J \cup J'} T_j = 1.64,\ C_{max} = 19.91$ )

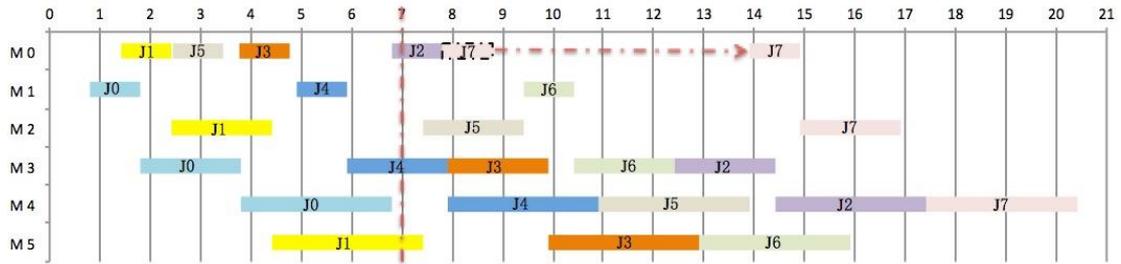

Fig. 8. The updated schedule of an optimized solution obtained by the proposed approach in a dynamic environment ($\sum_{j \in J \cup J'} T_j = 0.96, C_{max} = 20.42$)

4.4 Priority based GA operations on GPUs

- The fitness function: The parallel GA assesses the solutions based on the fitness function. In general, it is generated by the objective function to evaluate the solution domain. Since most shop scheduling problems are minimization problems [33] and the EDFFS is not an exception, the above-mentioned objective function (Eq. (1)) can be transformed into the fitness function as

Fitness funtion $= \max(E_{max} - (WT * \sum_{j \in J \cup J'} T_j + C_{max}), 0),$  (13)

where $E_{max}$ is the estimated maximum value of the objective function.

- The selection operation: On the basis of the value of fitness function, the larger fitness an individual has, the higher the chance it has to be chosen in the next generation. Because the 2D grid is adopted as the spatial population structure where each grid point contains one individual, the local asteroid selection is hired to make the selection operation. Moreover, since GPU texture caches are designed to gain an increase in performance accelerating access patterns with a great deal of spatial locality [34], we define the neighborhood on the grid always contains 5 individuals: the considered one and neighboring individuals as displayed in Fig, 9. Among these individuals, the tournament selection is implemented where the individual with the largest fitness value is the winner of each tournament and is selected to replace the considered individual.

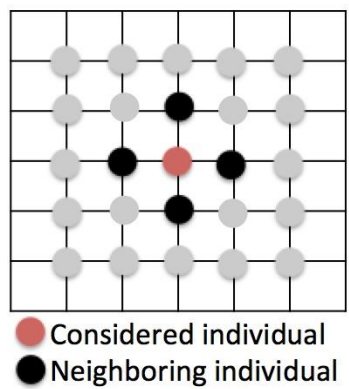

Fig. 9. The local asteroid selection

- The crossover operation: We pair individuals with neighbors (See Fig. 10.) rather than selecting two from population randomly. This strategy does not require global information sharing and is appreciated to work on 2D grid architecture. Meanwhile, a risk that it converges to the local minima can be eliminated by its cooperation with the local asteroid selection. In details, a 2D single point crossover is executed for the target machine matrix and the priority matrix respectively if a specified probability is satisfied. As the randomly generated values in the priority matrix is unique, a correction step is required to replace the duplicate values by the missing values in ascending order. An example shows the procedure in Fig. 11.

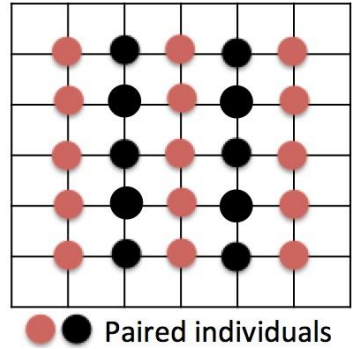

Fig.10. The neighboring paired crossover

Before crossover  ● $X(k) = \begin{bmatrix} -1 & 1 & 0 \\ -1 & -1 & 1 \\ 1 & 0 & 1 \end{bmatrix}, Y(k) = \begin{bmatrix} -1 & 6 & 1 \\ -1 & -1 & 5 \\ 4 & 3 & 2 \end{bmatrix}$  ● $X(k) = \begin{bmatrix} -1 & 0 & 0 \\ -1 & -1 & 0 \\ 1 & 1 & 1 \end{bmatrix}, Y(k) = \begin{bmatrix} -1 & 3 & 6 \\ -1 & -1 & 1 \\ 4 & 2 & 5 \end{bmatrix}$

After crossover  ● $X(k) = \begin{bmatrix} -1 & 1 & 0 \\ -1 & -1 & 0 \\ 1 & 1 & 1 \end{bmatrix}, Y(k) = \begin{bmatrix} -1 & 6 & 6 \\ -1 & -1 & 1 \\ 4 & 2 & 5 \end{bmatrix}$  ● $X(k) = \begin{bmatrix} -1 & 0 & 0 \\ -1 & -1 & 1 \\ 1 & 0 & 1 \end{bmatrix}, Y(k) = \begin{bmatrix} -1 & 3 & 1 \\ -1 & -1 & 5 \\ 4 & 3 & 2 \end{bmatrix}$

Correction  ● $X(k) = \begin{bmatrix} -1 & 1 & 0 \\ -1 & -1 & 0 \\ 1 & 1 & 1 \end{bmatrix}, Y(k) = \begin{bmatrix} -1 & 6 & 3 \\ -1 & -1 & 1 \\ 4 & 2 & 5 \end{bmatrix}$  ● $X(k) = \begin{bmatrix} -1 & 0 & 0 \\ -1 & -1 & 1 \\ 1 & 0 & 1 \end{bmatrix}, Y(k) = \begin{bmatrix} -1 & 3 & 1 \\ -1 & -1 & 5 \\ 4 & 6 & 2 \end{bmatrix}$

Fig.11. An example of the neighboring paired crossover

- The mutation operation：Any individual in the population gets a random number generated on the interval 0 to 1. If it is smaller than the default mutation rate, the mutation operation is executed in order to yield solutions with new information. The non-negative elements of the target machine matrix of this individual are replaced by random values in the range, apart from the original ones. Regarding the priority matrix, two non-negative elements are chosen randomly to exchange the values. An example is given in Fig. 12.

Before mutation  $X(k) = \begin{bmatrix} -1 & 1 & 0 \\ -1 & -1 & 1 \\ 1 & 0 & 1 \end{bmatrix}, Y(k) = \begin{bmatrix} -1 & 6 & 1 \\ -1 & -1 & 5 \\ 4 & 3 & 2 \end{bmatrix}$

After mutation  $X(k) = \begin{bmatrix} -1 & 0 & 1 \\ -1 & -1 & 0 \\ 0 & 1 & 0 \end{bmatrix}, Y(k) = \begin{bmatrix} -1 & 2 & 1 \\ -1 & -1 & 5 \\ 4 & 3 & 6 \end{bmatrix}$

Fig.12. An example of the mutation

- The replacement operation: the individual whose fitness is the largest in history within one island is kept. Then it is used to replace the individual whose fitness is the smallest within this island. As one island is presented as one CUDA block, this operation is carried through shared memory.

- The migration operation: Islands are interconnected as a single ring as shown in Fig. 13. An island can only accept an individual with the largest fitness value from one neighbor to overwrite the individual with the smallest fitness value. Shared memory is hired to search the best individual and the worst individual within one island while the overwriting is processed via global memory synchronously.

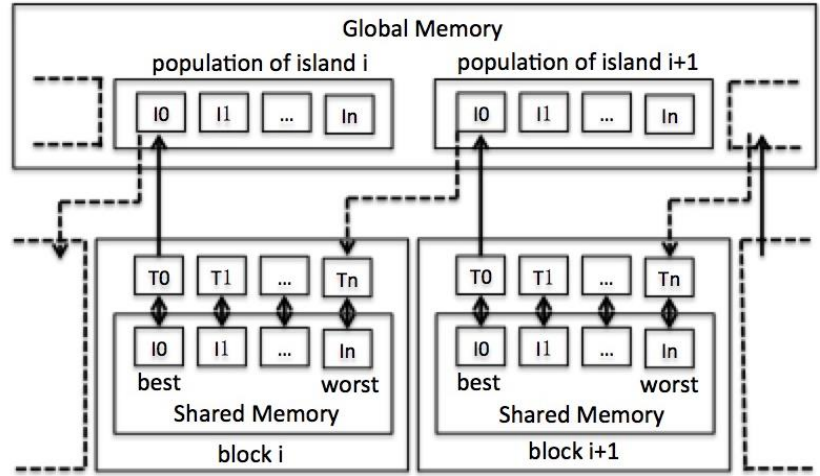

Fig. 13. The single ring migration among islands

## 5. Numerical experiments

To analyze the performance of the proposed algorithm, test 1 and test 2 are conducted in terms of an energy efficient FFS without considering new arrival jobs. Test 1 configures the parameters of the proposed hybrid GA, while test 2 shows its efficiency and effectiveness compared to the classical GA [29], the cellular GA [35] and the OpenMP based master-slave GA. New arrival jobs are included in test 3 to test the performance of the EDFFS. A small size instance is considered in those 3 tests. There are 10 original jobs with 3 production stages. Each stage includes 2 parallel machines. The power's peak is imposed through a bound equal to 4. Test 4 examines the convergence trend in the dynamic environment with 3 different size problems. The instances are characterized by the different numbers of jobs (n = 10, 50, 80) with the different numbers of stages (g = 3, 4, 4), the different numbers of machines (o = 2, 2, 3) in each stage and the different numbers of power's peak (Qmax =4, 5, 10). The rescheduling point is randomly generated in test 3 and test 4. The number of new arrival jobs is decided by the ratio of the rescheduling point on the makespan in the original schedule times the amount of original jobs. This is designed to keep the total amount of jobs waiting to be scheduled roughly consistent. The estimated maximum value of the objective function $E_{max}$ is set as $10^a$, where $a \in N_+$. The value of a is kept increasing from 1 until all individuals' initial objective function values are smaller than $E_{max}$. Other experimental relative data are defined in Table 5.

**Table 5**

The experimental relative data.

| | |
|---|---|
| WT | 100 |
| $P_{jsm}$ | U[1, 5], where $P_{0sm} = P_{1sm} = \cdots = P_{(n+n'-1)sm}$ |
| $R_j$ | U[0, $\bar{P}$], where $\bar{P} = \sum_s(\sum_m P_{jsm}/o)$ |
| $D_j$ | $R_j + \bar{P}(1 + \sigma)$, where $\sigma$=U[0,2] |
| $Q_{jsm}$ | 1 |

The experimental platform is based on Intel Xeon E5640 CPU with 2.67GHz clock speed. The GPU code implementation is carried out using CUDA 8.0 on a NVIDIA Tesla K40, with 2880 cores at 0.745GHz and 12 GB GDDR5 of global memory. All programs are written in C, except for the GPU kernels in CUDA C.

5.1 Parameters Configuration Test of the Hybrid Parallel GA

As the maximum threads amount per block on CUDA is 1024 and they are organized in grid, the maximum island size for the hybrid GA is 1024 (32×32). In order to have more than one island in all cases, the population size is kept as 6048 (64×64). Since small size islands with the migration lead to premature convergence while the algorithm with large size islands converges slower [31], we set there are 64 (8×8) individuals in one island. Furthermore, the values of crossover rate and mutation rate are given as 0.9 and 0.1 respectively on account of the existing experiences that the most appropriate crossover rate ranging between 0.75 and 0.9 [36] and the mutation rate is supposed to be much lower than the crossover rate [37].

In order to ensure the effectiveness of our GA parameters, we applied the parallel hybrid GA on the tested instance with three groups of crossover rates and three groups of mutation rates as in Table 6. According to the average results of 100 iterations, we could find the crossover rate and mutation rate do have some influence on the algorithm performance. Moreover, when crossover rate=0.9 and mutation rate=0.1, the parallel hybrid GA could obtain satisfying results on both solution quality and execution time. To achieve the fairness of comparison, we set the crossover rate and the mutation rate as 0.9 and 0.1 for all kinds of GAs in the following tests.

**Table 6**

Results of parallel hybrid GA on GPUs with different crossover rate and mutation rate settings (Generation Size =100)

| Crossover Rate | Mutation Rate | Solution Quality | Execution Time (s) |
| --- | --- | --- | --- |
| 0.75 | 0.05 | 216.39 | 8.21 |
| 0.75 | 0.1 | 219.98 | 8.34 |
| 0.75 | 0.15 | 211.70 | 8.47 |
| 0.825 | 0.05 | 220.39 | 8.30 |
| 0.825 | 0.1 | 214.03 | 8.43 |
| 0.825 | 0.15 | 210.90 | 8.53 |
| 0.9 | 0.05 | 216.56 | 8.36 |
| 0.9 | 0.1 | 209.81 | 8.50 |
| 0.9 | 0.15 | 215.09 | 8.58 |

Due the influence from the island size, the trend of the probability obtaining adequate solutions with different island sizes is illustrated in Fig.14. Each point in the figure denotes the rate over 100 runs. Regarding the values of the objective function got by different groups of crossover rates and mutation rates after 100 generations are approaching to 200, we set the adequate solution level as 200 for the tested instance. When a value of the objective function is less than 200 after the specified generations, it is considered as an adequate solution. From Fig.14 and Table 7 we could observe a great influence from the island size on the solutions' quality of the hybrid parallel GA while but a few difference on the execution time. The islands with 64 individuals (8×8 threads) perform best. In terms of the 2D population size 4096 (64×64), there are 64 islands (8×8 blocks).

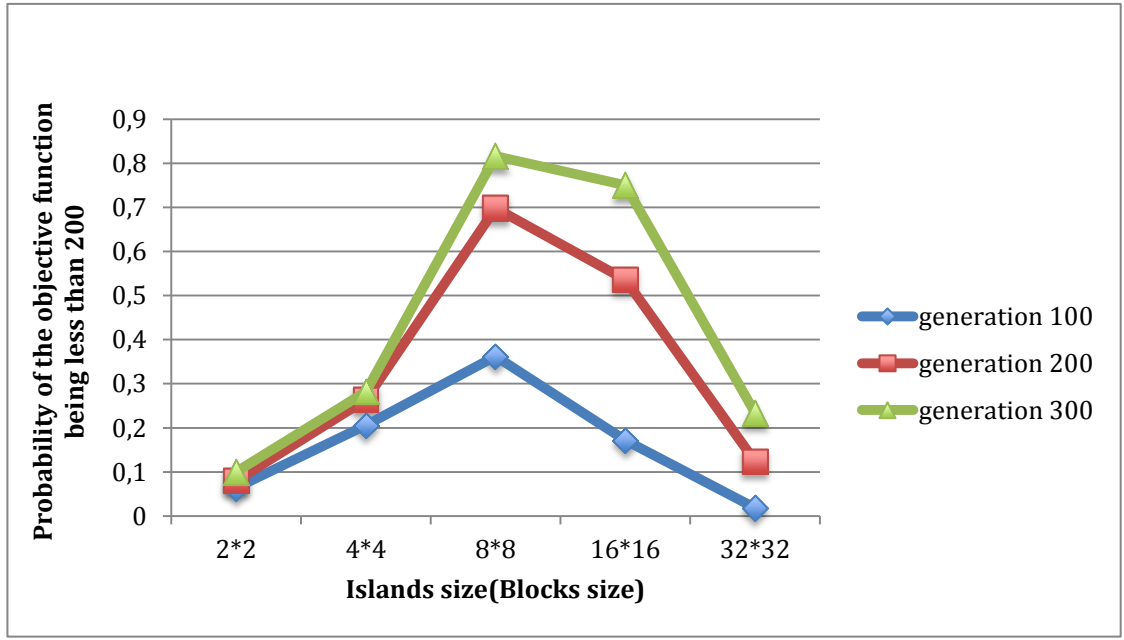

Fig. 14. The trend of the probability obtaining adequate solutions with different island sizes (block sizes) on GPUs

**Table 7**

Execution Time with different island sizes (block sizes) on GPUs (s)

| Generations \ Island Size | 4 (2×2) | 16 (4×4) | 64 (8×8) | 256 (16×16) | 1024 (32×32) |
|---|---|---|---|---|---|
| 100 | 7.65 | 7.71 | 9.11 | 9.14 | 12.30 |

5.2 Performance Evaluation Test of the Hybrid Parallel GA

Firstly, we try to compare the solutions obtained from the hybrid parallel GA, the classical GA and the cellular GA. As the roulette wheel selection is the most frequently used selection strategy [38], we take it for the classical GA while the single point crossover is executed with randomly paired individuals. Meanwhile, the mutation operation is kept the same as the hybrid parallel GA. The cellular GA is a popular way to apply the conventional GA in grid environments and has been implemented a lot to solve combinatorial optimization problems [13, 39, 40]. In this case, two individuals are selected from a similar neighborhood area as the local asteroid selection. Then the single point crossover recombines the chromosomes from them to generate a new individual. Finally, the new individual takes the same mutation as other two GAs and replaces the target individual if its solution is better.

For fair comparison, an OpenMP based master-slave GA is also taken into consideration. The master-slave model exploits parallelism in the classical GA by distributing the most time consuming part, fitness function evaluation, to slaves. As it does not affect the behavior of the algorithm, the OpenMP based master-slave GA is only included for execution time comparison. Moreover, we execute the hybrid parallel GA and the cellular GA with GPU, the classical GA with single core CPU, the OpenMP based master-slave GA with four cores CPU. Each of them is generated 100 times respectively.

**Table 8**

Solutions quality comparison

| Generations | Hybrid Parallel GA | | Classical GA | | Cellular GA | |
| --- | --- | --- | --- | --- | --- | --- |
| | Avg. | Best | Avg. | Best | Avg. | Best |
| 100 | 209.81 | 153.45 | 410.72 | 236.55 | 258.08 | 169.39 |
| 200 | 183.16 | 151.67 | 354.64 | 214.31 | 223.20 | 153.24 |
| 300 | 181.80 | 151.67 | 339.09 | 198.69 | 221.84 | 152.53 |
| 400 | 178.32 | 149.83 | 331.57 | 170.60 | 221.81 | 151.66 |
| 500 | 177.93 | 149.47 | 327.46 | 156.41 | 216.79 | 151.66 |

From the results in Table 8, we discover that the proposed hybrid parallel GA always gains better performance with the average value and the best value of the objective function than the classical GA and the cellular GA. Since, fine-grained models at the lower level could obtain good population diversity when dealing with high-dimensional variable spaces [17, 32] and island models at the upper level converge faster by subpopulations [31], the hybrid parallel GA combines the merits from both. Moreover, the cellular GA overcomes the classical GA as it allows a better exploration of the search space with respect to the decentralized population [13].

**Table 9**

Execution time comparison (Generations=100)

| Population size | Hybrid Parallel GA with GPUs (s) | Cellular GA with GPUs (s) | Classical GA with single core CPU (s) | OpenMP based Master-Slave GA with 4 cores CPU (s) |
|---|---|---|---|---|
| 64×64 | 8.77 | 8.14 | 129.16 | 39.50 |
| 128×128 | 30.71 | 31.13 | 554.01 | 182.27 |
| 256×256 | 105.73 | 108.07 | 2651.61 | 1127.78 |

Since the hybrid GA and the cellular GA are designed specially for 2 dimensional grid architectures, they could combine the benefits from the CUDA framework and almost take the same execution time when dealing with different population sizes as illustrated in Table 9. On the opposite, the classical GA with single core CPU takes from 14.73 times to 25.08 times execution time of the hybrid parallel GA when the population size is increased from 64×64 to 256×256. With the development of multi-cores CPU, the OpenMP based master-slave GA improves the performance a lot by distributing the fitness function evaluation to slaves and executing them concurrently. However, due to the limited amount of threads, the speedup from 4 cores CPU is not as significant as the one from GPUs, especially with the large population size. Moreover, we expect the hybrid parallel GA can achieve even further acceleration for complicated or larger-scale problems that require huge population size by its fully implemented parallelism with GPUs.

5.3 Sensitive Analysis Test of the EDFFS

As the number of new arrival jobs is equal to the ratio RS to the tmakespan in the original schedule) that multiplies the amount of original jobs, we change the amount of new arrival jobs by varying the ratio of the RS to the makespan in the original schedule. The influence with different ratios to the predictive reactive complete rescheduling approach and the traditional static approach are displayed in Table 10. The iteration number is kept as 100 like the last two tests. The predictive reactive complete rescheduling approach is more flexible in a dynamic environment as it reschedules the new arrival jobs at the beginning of the rescheduling point. However, those jobs could only be scheduled after completing the operations of the original schedule at each stage by the traditional static approach. This impact is more evident when the ratio of the RS to the makespan in the original schedule is small. And it is decreasing and almost disappears when the RS takes place near the end of the original schedule. Therefore, we strongly suggest using the predictive reactive complete rescheduling approach with the assistance of GPUs when the RS is arranged at the first half part of the original schedule. Meanwhile, the traditional static approach may have similar performance if the RS is considered at later half part.

**Table 10**

The comparison between the predictive reactive complete rescheduling approach and the traditional static approach with different ratios of the RS to the makespan in the original schedule (Generations=100)

| Ratio of the RS to the makespan in the original schedule | Traditional static approach | Predictive reactive complete rescheduling approach | Improvement Ratio |
|---:|---:|---:|---:|
| 20% | 2727.61 | 1195.29 | 2.28 |
| 40% | 5608.31 | 3745.06 | 1.50 |
| 60% | 5464.86 | 4673.57 | 1.17 |
| 80% | 4805.43 | 4755.91 | 1.01 |

As tardy jobs typically cause penalty costs [28] and have a great influence on customer satisfaction, the weight WT indicates the priority of the total tardiness in the objective function. However, we consider the relationship between two objectives with different WT settings due to the importance of makespan in manufacturing practice and Table 11 shows the average results of 100 iterations. According to the values of total tardiness and makespan, we could find the makespan is less sensitive to the weight of WT than the total tardiness as the variance of makespan is 0.46 while the variance of total tardiness is 33.53. Thus, manufacturers should take the chance to optimize the total tardiness while limiting the makespan in a reasonable range.

**Table 11**

Relationship between two objectives with different WT settings (Generations=100)

| WT | Total Tardiness | Makespan | Objective Function Value |
|---|---|---|---|
| 0.01 | 35.23 | 40.54 | 40.89 |
| 0.1 | 23.43 | 40.78 | 43.12 |
| 0.4 | 19.04 | 41.14 | 48.76 |
| 0.7 | 18.61 | 41.23 | 54.26 |
| 1 | 18.29 | 41.44 | 59.73 |
| 4 | 17.83 | 42.15 | 113.46 |
| 7 | 17.69 | 42.18 | 166.00 |
| 10 | 17.57 | 42.12 | 217.86 |
| 100 | 17.58 | 42.39 | 1800.43 |
| Variance | 33.53 | 0.46 | |

5.4 Convergence trend test of the EDFFS

As a GA converges when most of the population is identical or the diversity is minimal [41], there is no need to execute the algorithm for more generations after the convergence point. For the EDFFS, it is important to identify the convergence point and its corresponding execution time for different size problems. Three different size problems are considered in this test. The convergence trends of the small size, the medium size and the large size problem instances are described in Fig. 15, Fig. 16, Fig 17 separately. Each point in figures displays the average value of 30 runs.

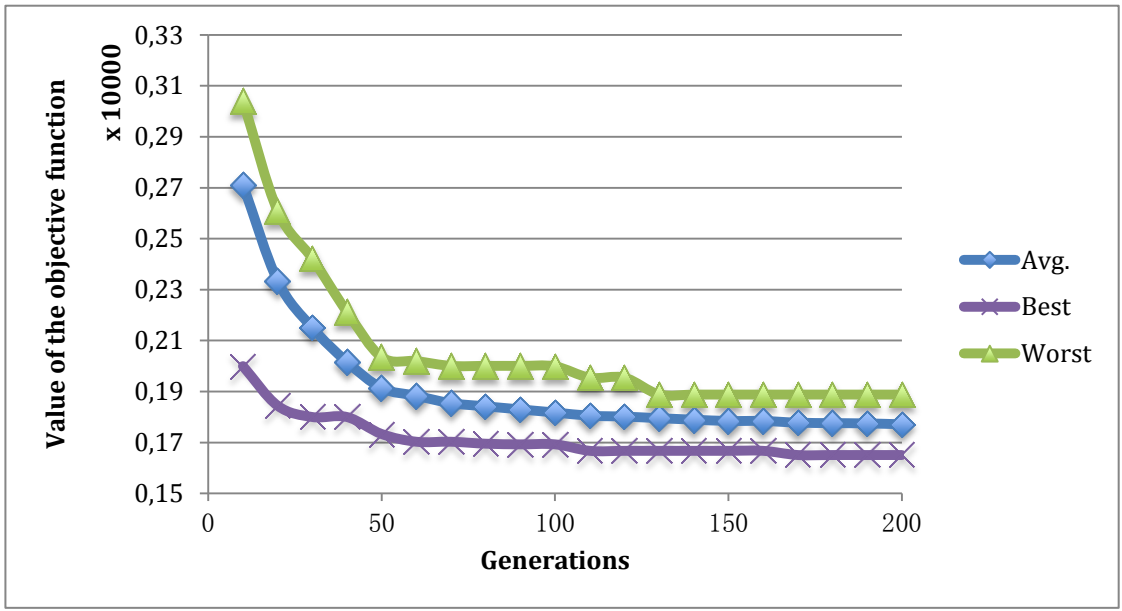

Fig. 15. The convergence trend of the small size problem

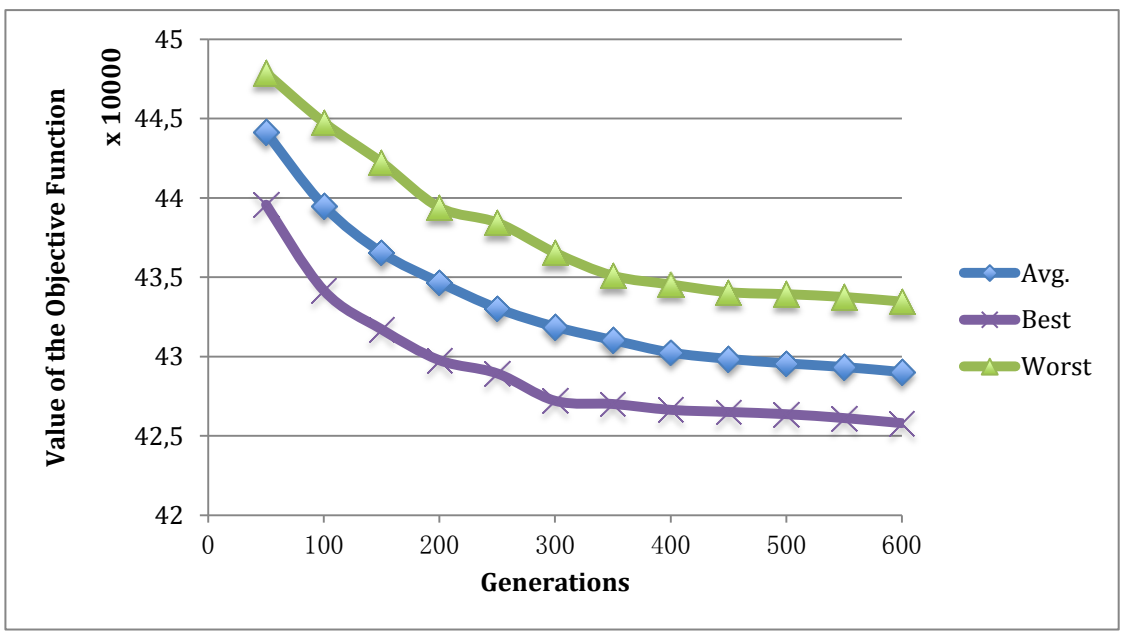

Fig. 16. The convergence trend of the medium size problem

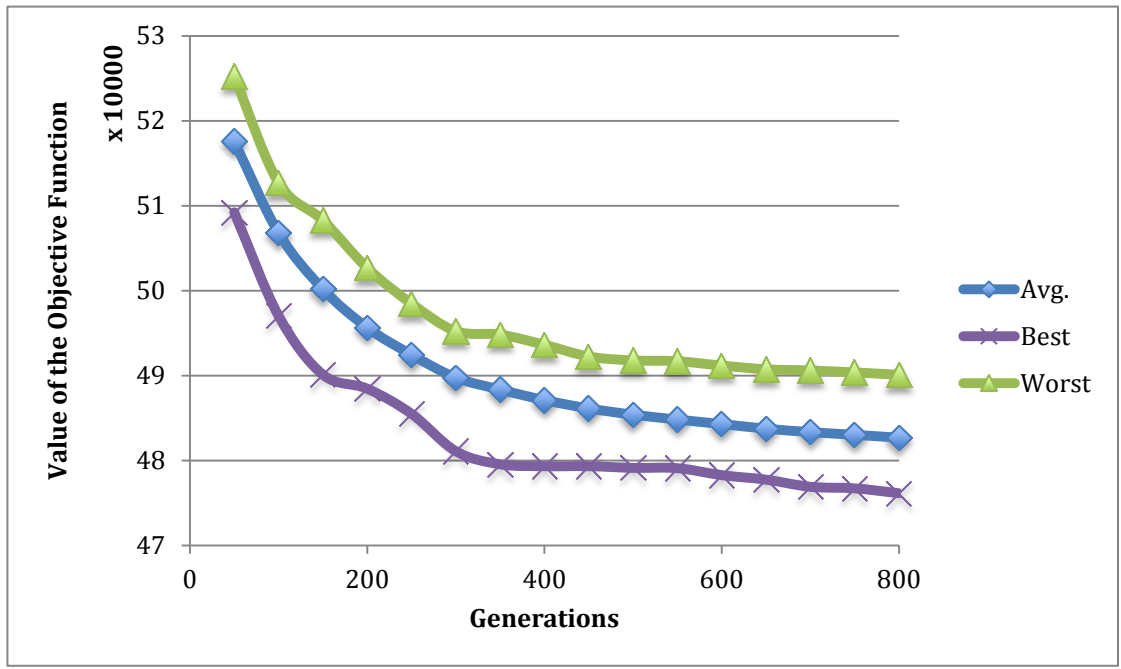

Fig. 17. The convergence trend of the large size problem

With regard to the small size problem, it converges approximately at the level of 50 generations, while the value for the medium size and the large size problems is around 400 and 500. As the complexity increases when we raise the size of the problem, the execution time per 10 generations for those problems is about 1.24s, 223.37s and 4256.14s respectively. Therefore, to get solutions after convergence for the small size problem, it takes 6.2s whereas the medium size and the large size problems need much longer time as 8934.8s and 212807s. Due to the dramatically increasing execution time for large-scale problems, the parallel GA may get a feasible solution before achieving the convergence based on decision-makers' consideration, namely a trade-off between the solution quality and the time consumption.

## 6. Conclusions

In this paper, we have first studied a dynamic energy efficient flexible flow shop scheduling model using peak power value with the consideration of new arrival jobs. To solve this NP-hard problem, a priority based hybrid parallel GA with a predictive reactive complete rescheduling approach was developed. In order to have a short response in the dynamic environment, we proposed a parallel GA. It consists of a fine-grained GA at the lower level and an island GA at the upper level. This parallel GA is highly consistent with the hierarchy of threads and different types of memory of CUDA framework. In the first test, a discussion was conducted to obtain the reasonable island size and island amount for the related problem by inhibiting the premature convergence with a faster convergence speed. Afterwards, our method displayed in test 2 showed that it could gain better results than the classical GA on CPU through the advantages from both the fine-grained GA and the island GA. But it also reduces the time requirements dramatically. Moreover as seen in test 3, the proposed approach has better performance than the traditional static approach because of its flexibility. Finally, test 4 demonstrated the response time to achieve the convergence point for large-scale problems. We suggest as well in this case decision-makers to obtain a feasible scheduling by making a trade-off between the solution quality and the time consumption.

# Acknowledgement

This work was supported by a scholarship from the China Scholarship Council (CSC). Moreover, The authors would like to express their gratitude to Bastien Plazolles for his helpful comments.